\documentclass [a4paper,fleqn]{article}

\usepackage {amsmath} \usepackage{amssymb} \usepackage{cite} \usepackage{amsthm}
\numberwithin{equation}{section} \numberwithin{table}{section} \mathindent=0pt
\theoremstyle{plain} \newtheorem{theorem}{Theorem}

\numberwithin{theorem}{section}

\begin{document}

\title{Nonlinear differential equations \\with exact solutions}
\author{N.A. Kudryashov}
\date{Department of Applied Mathematics\\
Moscow  Engineering and Physics Institute\\
(State university)\\
31 Kashirskoe Shosse,  115409\\
Moscow, Russian Federation} \maketitle

\begin{abstract}
New problem is considered that is to find nonlinear differential equations with special
solutions. Method is presented to construct nonlinear ordinary differential equations
with exact solution. Crucial step to the method is the assumption that nonlinear
differential equations have exact solution which is general solution of the simplest
integrable equation. The Riccati equation is shown to be a building block to find a lot
of nonlinear differential equations with exact solutions. Nonlinear differential
equations of the second, third and fourth order with special solutions are given. Most of
these equations are used at the description of processes in physics and in theory of
nonlinear waves.
\end{abstract}

\emph{Keywords:} Riccati equation, nonlinear differential equation, exact solution\\

PACS: 02.30.Hq - Ordinary differential equations

\section{Introduction}
In recent years one can observe a splash of papers where authors presented a lot of
different approaches to look for exact solutions of nonlinear differential equations
\cite{1,2,3,4,5,6,7,8,9}. There are two reasons to make the study in this direction.
First there is a great interest to the investigation of nonlinear phenomenon. Secondly we
have codes Maple, Mathematica and other ones to conduct a lot of symbolical calculations.

It is well known that all nonlinear differential equations can be connectionally divided
into three types: exactly solvable, partially solvable and those that have no exact
solution.

Consider nonlinear evolution equation

\begin{equation}
\label{1.1}E_1[u] \equiv E_1(u, u_t, u_x, ..., x, t)=0
\end{equation}

Assume we need to have exact solutions of this equation. First of all one can try to
solve a question about the integrability of this equation. For this aim one can apply the
Painlev\'e test to check the integrability of this equation \cite{10}. As a result we can
have two variants. First variant is the equation \eqref{1.1} pass the Paileve\'e test and
we have the necessary condition for integrability of equation in this case. Second
variant is the equation \eqref{1.1} does not pass the Painlev\'e test. In this case we
can look for some transformation to obtain equation that can be passed the Painlev\'e
test. However we often obtain that the equation \eqref{1.1} does not pass the Painlev\'e
test and we have not got any transformation to obtain good form for the origin equation.
Unfortunately a lot of nonlinear evolution equation does not pass the Painlev\'e test. In
any case one can search exact solutions of nonlinear differential equation. In fact one
can look for exact solutions of nonlinear differential equation of all types without
solution of the problem about the integrability.

Usually we look for exact solution of nonlinear evolution equation
taking into account the travelling wave and search exact solution
of equation \eqref{1.1} in the form

\begin{equation}
\label{1.2}u(x, t) = y(z), \qquad z=x-C_0 t
\end{equation}

As a result we have that the equation \eqref{1.1} reduces to the nonlinear ordinary
differential equation (ODE)

\begin{equation}
\label{1.3}E_2[y] \equiv E_2(y, y_z, ...,z)=0
\end{equation}

To obtain exact solutions of equation \eqref{1.3} one can apply
different approaches \cite{2,3,11,12,13,14,15,16,17,18,19,20}.
However one can note that the most methods that are used to search
exact solutions take into account the singular analysis for
solutions of the nonlinear differential equations
\cite{10,11,12,21,22,23}. Even though the investigators are not
aware of the analytical theory of nonlinear differential equations
they apply as a rule its ideas and approaches.

Using the singular analysis first of all we have to consider the
leading members of equation \eqref{1.3}. After that we find the
singularity for solution of equation \eqref{1.3}. Further the
truncated expansion is used to have the transformation to search
exact solutions of nonlinear ODEs. At this point one can use some
trial functions (hyperbolic and elliptic and so on) to look for
exact solutions of nonlinear ODEs \cite{2,3,24,25,26,27,28,29,30}.

However one can note that hyperbolic and elliptic functions are
general solutions of nonlinear exactly solvable equations. We have
as a rule that partially solvable nonlinear differential equations
have exact solutions that are general solutions of solvable
equations of lesser order.

In this paper we are going to start the solution of the new problem that is to find
nonlinear ordinary differential equations of polynomial form which have special
solutions. Let us explain the idea of our work. It is well known there is the great
problem that is to find nonlinear differential equations that are integrable. It is very
important because we often want to have general solution of nonlinear ordinary equations.
For a example Painlev\'e and his school found 50 canonical forms of the second order
class for the ordinary differential equations that are integrable equations. However
sometimes we can content ourselves with some special solutions because a lot of
differential equations are nonintegrable ones although they are intensively used in
physics and look as simple equations. It is important to find some special solutions of
these equations that are called exact solutions. In this connection it is convenient to
have the list of nonlinear differential equations that are not integrable but have
special solutions.

The aim of this work is to present new method to find nonlinear differential equations
with exact solutions. Using our approach we give a class of nonlinear ordinary
differential equations that have exact solutions. Special solutions of our nonlinear
ordinary differential equations are found via general solution of the Riccati equation.

As this takes place we do not solve the problem of finding all possible exact solutions
for our class of nonlinear ordinary differential equations. Our class contains a lot of
exactly solvable equations. In this paper we give only the polynomial class of nonlinear
ordinary differential equations with special solutions.

The outline of this paper is as follows. In section 2 we present the method to find
nonlinear ordinary differential equations (ODEs) with special solutions. Nonlinear ODEs
with exact solutions of the first, second, third and fourth  degree singularities are
given in section 3, 4, 5 and 6. Example of nonlinear ODE with exact solution of the fifth
degree singularity  is considered in section 7.

\section{Method applied}

Let us discuss the method that can be applied to find nonlinear differential equations
with exact solutions. One can note that most nonlinear ordinary differential equations
has exact solutions that are general solutions of differential equations of lesser order.
Much more than that exact equations for the most nonlinear differential equations are
determined via general solution of the Riccati equation.  This is so indeed because one
can note that the most approaches to search exact solutions of nonlinear ordinary
differential equations are based on general solution of the Riccati equation. The
application of the tanh method confirms this idea \cite{24,28,29,31,32}. This is due to
the fact that the Riccati equation is simplest equation of first order of polynomial form
that have the Painleve property \cite{10,17,21}.

The Riccati equation can be presented in the form

\begin{equation}
\label{2.1}R[Y]=Y_z + Y^2 + q(z) =0
\end{equation}

We have the following simple theorem.

\begin{theorem}
\label{T:2.1.} Let $Y(z)$ be solution of equation \eqref{2.1} than equations

\begin{equation}
\label{2.2}Y_{zz} = 2Y^3 + 2qY - q_z
\end{equation}

\begin{equation}
\label{2.3}Y_{zzz} =-6Y^4 -8qY^2 + 2 q_z Y - 2 q^2 -q_{zz}
\end{equation}

\begin{equation}
\label{2.4}Y_{zzzz} =24Y^5 + 40 q Y^3 + 16 q^2 Y - 10 q_z Y^2 + 2 q_{zz} Y - 6 qq_z -
q_{zzz}
\end{equation}
have special solutions that are expressed via general solution of equation \eqref{2.1}
\end{theorem}

\begin{proof}Theorem 2.1 is proved by differentiation of \eqref{2.1} with respect to $z$
and substitution $Y_z$ from equation \eqref{2.1} and so on into expressions obtained.

\end{proof}
\textit{Corollary 2.1.} Equations

\begin{equation}
\label{2.5}Y_{zz}= 2Y^3 - 2\alpha Y
\end{equation}

\begin{equation}
\label{2.6}Y_{zzz} =- 6Y^4 + 8\alpha Y^2 - 2\alpha^2
\end{equation}

\begin{equation}
\label{2.7}Y_{zzzz} =24 Y^5 - 40 \alpha Y^3 +16 \alpha^2 Y
\end{equation}
have special solutions $Y(z)$ where $Y(z)$ is the general solutions of the Riccati
equation in the form

\begin{equation}
\label{2.8}Y_{z}+ Y^2 - \alpha =0
\end{equation}

It is well known that general solutions of equation \eqref{2.8} is

\begin{equation}
\label{2.13}Y(z) = \sqrt{\alpha} \tanh \left(\sqrt{\alpha} z + \varphi_0\right)
\end{equation}
where $\varphi_0$ is arbitrary constant.

We want to find nonlinear ordinary differential equation which
have special solutions that are determined via general solution of
the Riccati equation.

Algorithm of our method can be presented by \textit{four steps}. At \textit{the first
step} we choose the singularity of special solution and give the form of this solution.
At \textit{the second step} we take the order of nonlinear ordinary differential equation
what we want to search. \textit{The third step} lies in the fact that we write the
general form of nonlinear differential equation taking into account the singularity of
the solution and the given order for nonlinear differential equation. \textit{The fourth
step} contains calculations. As a result we find limitations for the parameters in order
for nonlinear differential equation has exact solutions. At this step we have nonlinear
ODE with exact solutions.

Let us demonstrate our approach. With this aim let us find a
nonlinear ordinary differential equation of the second order with
solution of the second degree singularity. This solution takes the
form

\begin{equation}
\label{2.16}y(z) = A_0 + A_1 Y + A_2 Y^2
\end{equation}
Here $Y(z)$ is a solution of equation \eqref{2.8}. Without loss of generality we can take
 $A_2=-12$.

First of all let us write the general form of the nonlinear second order ordinary
differential equation with solution \eqref{2.16}. It takes the form

\begin{equation}
\label{2.17}E_1 \equiv  y_{zz} + a_0 y^2 - b_0 y_z -C_0 y + C_1 =0
\end{equation}

Here $a_0, b_0,  C_0$ and $C_1$ are unknown parameters of equation

This equation was written using singularity of solution
\eqref{2.16}. Actually the singularity of the first term in
\eqref{2.17} is equal to 4 and for the polynomial form of the
equation we have to take into account the same singularity of
nonlinear expression for the second term \eqref{2.17}. Other terms
in \eqref{2.17} have lesser singularity.

We want to find equation \eqref{2.17} that has exact solution
\eqref{2.13}. We can do this if we define values of parameters
$a_0, b_0, C_0$ and $C_1$. Substituting \eqref{2.16} at $A_2=-12$
into equation \eqref{2.17} and taking into account equations
\eqref{2.5} and \eqref{2.8} we have equation in the form

\begin{equation}
\begin{gathered}
\label{2.18}\left( -72+144\,a_{{0}} \right)   Y ^ {4}+ \left(
-24\,a_{{0}}A_{{1}}+2\,A_{{1}}-24\,b_{{0}} \right)
  Y  ^{3}+\\
 \\
 + \left( -24\,a_{{0}}A_{{0}}+a
_{{0}}{A_{{1}}}^{2}+12\,C_{{0}}+b_{{0}}A_{{1}}+96\,\alpha \right)
  Y ^{2}+\\
 \\
 + \left( -C_{{0}}A_{{1}}-2\,A_
{{1}}\alpha+2\,a_{{0}}A_{{0}}A_{{1}}+24\,b_{{0}}\alpha \right) Y-
\\
 \\
  -24\,{\alpha}^{2}+a_{{0}}{A_{{0}}}^{2}-C_{{0}}A_{{0}
}-b_{{0}}A_{{1}}\alpha+C_{{1}}=0
\end{gathered}
\end{equation}

Equating expressions in equation \eqref{2.18} at different degrees of $Y$ to zero we have
algebraic equations for the parameters  $a_0, a_1, b_0$ and $C_1$ in the form

\begin{equation}
\label{2.19v}a_0 =\frac12
\end{equation}

\begin{equation}
\label{2.19}A_{{1}}=-{\frac {12}{5}}\,b_{{0}}
\end{equation}

\begin{equation}
\label{2.20}A_{{0}}=\frac1{25}\,{b_{{0}}}^{2}+C_{{0}}+8\,\alpha
\end{equation}

\begin{equation}
\label{2.21}\alpha={\frac {1}{100}}\,{b_{{0}}}^{2}
\end{equation}

\begin{equation}
\label{2.22}C_{{1}}=-{\frac {18}{625}}\,{b_{{0}}}^{4}+\frac12{C_{{0}}}^{2}
\end{equation}

As a result we obtain the nonlinear second order differential equation in the form

\begin{equation}
\label{2.24}y_{zz} +\frac12 y^2 - b_{{0}} y_z -C_{{0 }}y  -{\frac
{18}{625}}\,{b_{{0}}}^{4}+\frac12{C_{{0}}}^ {2}=0
\end{equation}

Exact solution of equation \eqref{2.24} is determined by the formula

\begin{equation}
\label{2.25}y \left( z \right) ={\frac {3}{25}}\,{b_{{0}}}^{2}+C_{{0}}-{\frac {12}
{5}}\,b_{{0}}Y  -12 Y ^{2}
\end{equation}

Where $Y(z)$ is a solution of the Riccati equation

\begin{equation}
\label{2.26}Y_z =-Y^2 +\frac{b_0^2}{100}
\end{equation}
which is

\begin{equation}
\label{2.26v}Y(z) = \pm \frac{b_0}{10} \tanh \left(\pm \frac{b_0 z}{10} +
\varphi_0\right)
\end{equation}

Here $\varphi_0$ is arbitrary constant. Actually we find equation \eqref{2.24} with
fascinating history. The matter is this equation can be found from the Korteveg -- de
Vries -- Burgers equation which takes the form

\begin{equation}
\label{2.27}u_t +uu_x + u_{xxx} =  b_0 u_{xx}
\end{equation}

Equation \eqref{2.27} is a generalization of the Korteveg -- de
Vries equation in the case of dissipative processes \cite{33}.
Equation \eqref{2.27} in contradistinction to the Korteveg -- de
Vries equation is not integrable equation.

Using the travelling wave \eqref{1.2} we have from \eqref{2.27} after integration over
$z$ the following equation

\begin{equation}
\label{2.28}C_1 -C_0 y +\frac12 y^2 + y_{zz} -  b_0y_z =0
\end{equation}

We can see that equation \eqref{2.28} is equation \eqref{2.24} as well.

However exact solutions of equation \eqref{2.28}were unknown over
many years. We hope that exact solutions \eqref{2.25} of equation
\eqref{2.22} were found first in the work \cite{11}. However this
exact solution have rediscovered in a number of papers
\cite{34,35,36}.

Assuming $b_0=0$ in equation \eqref{2.24} we have exactly solvable
equation solution. This one can be found via the elliptic
function. Nonlinear evolution equation \eqref{2.27} in this case
takes the form of the Korteveg -- de Vries equation and solution
\eqref{2.25} is the soliton.

One can see we can obtain  nonlinear integrable ODEs too using our
approach.

We have to note that equation \eqref{2.24} can be transformed to
nonlinear ODE that is found from the nonlinear evolution equation

\begin{equation}
\label{2.29}u_t = u_{xx} -  C_0 u - d_0 u^2
\end{equation}

This is the Fisher equation \cite{37} or the Kolmogorov --
Petrovskii -- Piskunov equation \cite{38}. Special solutions of
this equation were found first in the work \cite{39} and repeated
a lot of times later.

\section{Nonlinear ODEs with exact solutions of the first degree singularity}

Taking into account nonlinear ODEs \eqref{2.5} -- \eqref{2.8} one can find nonlinear
differential equations with exact solutions. These exact solutions are expressed via the
general solution of the Riccati equation

\begin{equation}
\label{3.1} Y_z =-Y^2 +\alpha
\end{equation}

First of all let us assume that nonlinear ODEs have special solutions of the first degree
singularity

\begin{equation}
\label{3.2} y(z) = A_0 + A_1 Y(z)
\end{equation}

 where $Y(z)$ is a
solution of the equation \eqref{3.1}.

Without loss of generality we can assume $A_0=0$ and $A_1=1$.

\textit{\textbf{3.1. Second order ODEs.}} Let us find the nonlinear second order ODEs
with exact solutions \eqref{3.2} where $Y(z)$ is a solution of equation \eqref{3.1}.

Second order ODEs with solutions of the first degree singularity
can take the form

\begin{equation}
\begin{gathered}
\label{3.3}y_{zz} + a_0 yy_z + a_1 y^3 - b_0 y_z + b_1 y^2 - \alpha C_0 y + \alpha C_1=0
\end{gathered}
\end{equation}

Here $a_0,\,\, a_1,\,\, b_0,\,\, b_1,\,\, C_0$ and $C_1$ are unknown coefficients of
equation \eqref{3.3}. Our problem is to find some relations between coefficients of
equation \eqref{3.3} so that equation \eqref{3.3} has exact solution \eqref{3.2}.

Substitution of \eqref{3.2} and Eqs. \eqref{3.1} and \eqref{2.5} into equation
\eqref{3.3} lead to relation

\begin{equation}
\begin{gathered}
\label{3.4}(a_1 -a_0 +2 ) Y^3 + ( b_1 + b_0 ) Y^2 + \alpha (a_0 - C_0 -2 ) Y + \alpha C_1
- \alpha b_0 =0
\end{gathered}
\end{equation}

From equation \eqref{3.4} we have

\begin{equation}
\begin{gathered}
\label{3.5}a_1 = a_0 -2,\\
\\b_1 = -b_0,\\
\\a_0 =C_0 + 2,\\
\\b_0 =C_1
\end{gathered}
\end{equation}

As a result we obtain the second order ODEs with exact solutions \eqref{3.2} at $A_0=0$
and $A_1=1$ in the form

\begin{equation}
\label{3.6}y_{zz} + \left((2 + C_0) y- C_1\right)y_z+ C_0 y^3 - C_1 y^2 -\alpha C_0 y +
\alpha C_1=0
\end{equation}

Here $\alpha$ is arbitrary constant.

At $C_0=-2$ we get ODE

\begin{equation}
\label{3.7}y_{zz}- C_1 y_z - 2 y^3 - C_1 y^2 + 2\alpha y + \alpha C_1 =0
\end{equation}

This equation can be found from the modified Korteveg -- de vries -- Burgers
equation

\begin{equation}
\label{3.8} u_t - \left(2 C_1 u + 6 u^2\right)u_x+ u_{xxx} = C_1 u_{xx}
\end{equation}
using the travelling wave \eqref{1.2}. We have found that equation
\eqref{3.8} has solution \eqref{3.2}. Exact solutions of equation
was found in work \cite{40}.

\textit{\textbf{3.2. Third order ODEs}}. Now let us find the nonlinear third order
ordinary differential equations with exact solutions \eqref{3.2}.

Nonlinear third order ordinary differential equations of the
polynomial type can have exact solution of the first degree
singularity if this one takes the form

\begin{equation}
\begin{gathered}
\label{3.2v} a_{{0}}y_{{{\it zzz}}}+a_{{1}}yy_{{{\it zz}}}+a_{{2}}{y}^{2}y_{{z}}+a_
{{3}}{y_{{z}}}^{2}+a_{{4}}{y}^{4}+b_{{0}}y_{{{\it zz}}}+b_{{1}}yy_{{z}
}+b_{{2}}{y}^{3}+\\
\\+d_{{0}}y_{{z}}+d_{{1}}{y}^{2}-\alpha\,{ C_0}\,y+ \alpha\,{ C_1}=0
\end{gathered}
\end{equation}

Substitution \eqref{3.2}, \eqref{3.1}, \eqref{2.5} and \eqref{2.6}
into equation \eqref{3.2v} leads to the equation

\begin{equation}
\begin{gathered}
\label{3.3v} \left( a_{{4}}-a_{{2}}+a_{{3}}-6\,a_{{0}}+2\,a_{{1}} \right)   Y ^{4}+
\left( 2\,b_{{0}}+b_{{2}}-b_{{1}}
 \right) {Y}^{3}+\\
 \\
 + \left( -a_{{2}}\alpha+d_{{1}}-d_{{0}}+2\,a_{{3}}
\alpha+2\,a_{{1}}\alpha-8\,a_{{0}}\alpha \right) {Y}^{2}+ \\
\\
+ \left( 2\,b_ {{0}}\alpha-b_{{1}}\alpha-\alpha\,{ C_0} \right) Y+\alpha\,{ C_1}-
2\,a_{{0}}{\alpha}^{2}-d_{{0}}\alpha+a_{{3}}{\alpha}^{2}=0
\end{gathered}
\end{equation}

We have from equation \eqref{3.3v}

\begin{equation}
\begin{gathered}
\label{3.4v}a_4 =a_2 + 6 a_0 -a_3 -2 a_1
\end{gathered}
\end{equation}

\begin{equation}
\begin{gathered}
\label{3.5v}b_2 = -C_0
\end{gathered}
\end{equation}

\begin{equation}
\begin{gathered}
\label{3.6v}d_1 = d_0 + \alpha (a_2 -2a_3 + 8a_0 -2a_1)=0
\end{gathered}
\end{equation}

\begin{equation}
\begin{gathered}
\label{3.7v}b_1 =2b_0 -C_0=0
\end{gathered}
\end{equation}

\begin{equation}
\begin{gathered}
\label{3.8v}\alpha=\frac{d_0 -C_1}{a_1-2a_0}
\end{gathered}
\end{equation}

As a result we obtain the nonlinear third order differential equation in the form

\begin{equation}
\begin{gathered}
\label{3.9v}a_{{0}}y_{{{\it zzz}}}+ \left( a_{{1}}y+b_{{0}} \right) y_{{{\it zz}}}
+a_{{3}}{y_{{z}}}^{2}+ \left( a_{{2}}{y}^{2}+2\,yb_{{0}}-yC_{{0}}+d_{{0 }} \right)
y_{{z}}+\\
\\
+ \left( a_{{2}}-a_{{3}}+6\,a_{{0}}-2\,a_{{1}}
 \right) {y}^{4}-C_{{0}}{y}^{3}+\\
 \\
 + \left( a_{{2}}\alpha-2\,a_{{3}}\alpha
+d_{{0}}+8\,a_{{0}}\alpha-2\,a_{{1}}\alpha \right) {y}^{2}-\alpha\,C_{
{0}}y+\alpha\,C_{{1}}=0
\end{gathered}
\end{equation}

Solution of equation \eqref{3.9v} is found by formula \eqref{3.1} at $A_0=0$ and $A_1=1$.

Assuming $a_1=0,\,\,a_2=0,\,\,a_3=0,\,\, b_0=C_0/2$ in equation \eqref{3.9v} we get

\begin{equation}
\begin{gathered}
\label{3.10v} a_0y_{zzz} +\frac12 C_0 y_{zz} + d_0 y_z + 6a_0 y^4 -C_0 y^3 + (8\alpha a_0
+ d_0) y^2 - \alpha C_0 y + \alpha C_1=0
\end{gathered}
\end{equation}

This equation can be found from the generalized Kuramto -- Sivashinsky equation
\cite{11,17}

\begin{equation}
\begin{gathered}
\label{3.11v}u_t + \varepsilon u^3 u_x + \beta u_{xx} + \gamma u_{xxx} + \delta u_{xxxx}
=0
\end{gathered}
\end{equation}
if we look for solution of equation \eqref{3.11v} using the travelling wave \eqref{1.2}.
It this case we have from equation \eqref{3.11v}

\begin{equation}
\begin{gathered}
\label{3.12v}C_1 - C_0 y + \frac{\varepsilon}{4} y^4 + \beta y_z + \gamma y_{zz} +\delta
y_{zzz} =0
\end{gathered}
\end{equation}

One can see that equation \eqref{3.10v} and \eqref{3.11v} have exact solutions that are
expressed via general solution of the Riccati equation

\textit{\textbf{3.3. Fourth order ODEs.}} General form of
nonlinear ordinary differential equation of the fourth order can
be written as the following

\begin{equation}
\begin{gathered}
\label{3.2a} a_{{0}}y_{{{\it zzzz}}}+a_{{1}}yy_{{{\it zzz}}}+a_{{2}}{y}^{2}y_{{{ \it
zz}}}+a_{{3}}y_{{z}}y_{{{\it zz}}}+a_{{4}}y{y_{{z}}}^{2}+a_{{5}}{y
}^{3}y_{{z}}+a_{{6}}{y}^{5}+ \\
\\
+b_{{0}}y_{{{\it zzz}}}+b_{{1}}yy_{{{\it zz
}}}+b_{{2}}{y}^{2}y_{{z}}+b_{{3}}{y_{{z}}}^{2}+b_{{4}}{y}^{4}+h_{{0}}y _{{{\it
zz}}}+h_{{1}}yy_{{z}}+\\
\\
+h_{{2}}{y}^{3}+d_{{0}}y_{{z}}+d_{{1}}{y} ^{2} -C_0 \alpha y + C_1 \alpha=0
\end{gathered}
\end{equation}

Substituting equation \eqref{2.5} -- \eqref{2.8} into equation \eqref{3.2a} as a result
we have nonlinear fourth order ordinary differential equation with exact solutions that
  is determined via general solution of the Riccati equation \eqref{2.8}. It takes the form

\begin{equation}
\begin{gathered}
\label{3.21}a_{{0}}y_{{{\it zzzz}}}+ \left( a_{{1}}y+b_{{0}} \right) y_{{{\it zzz} }}+
\left( b_{{1}}y+a_{{3}}y_{{z}}+a_{{2}}{y}^{2}+h_{{0}} \right) y_{{ {\it zz}}}+\\
\\
+ \left( a_{{4}}y+b_{{3}} \right) {y_{{z}}}^{2}+\\
\\
+ \left( d_{{0 }}+a_{{5}}{y}^{3}+yC_{{0}}+b_{{2}}{y}^{2}-ya_{{4}}\alpha+2\,ya_{{1}}
\alpha-16\,ya_{{0}}\alpha \right)y_z+\\
\\
+\left(2\,ya_{{3}}\alpha+2y  h_{{0 }} \right) y_{{z}}+\left(
-a_{{4}}+2\,a_{{3}}+6\,a_{{1}}-24\,a_{{0}}- 2\,a_{{2}}+a_{{5}} \right)
{y}^{5}+\\
\\
+ \left( -b_{{3}}-2\,b_{{1}}+b_{{2}} +6\,b_{{0}} \right) {y}^{4}+\\
\\
+ \left( -a_{{5}}\alpha+a_{{4}}\alpha+2\,a_
{{2}}\alpha-6\,a_{{1}}\alpha-2\,a_{{3}}\alpha+24\,a_{{0}}\alpha+C_{{0} } \right)
{y}^{3}+\\
\\
+ \left( 2\,b_{{3}}\alpha-b_{{2}}\alpha-8\,b_{{0}} \alpha+d_{{0}}+2\,b_{{1}}\alpha
\right) {y}^{2}-C_{{0}}\alpha\,y+C_{{1 }}\alpha=0
\end{gathered}
\end{equation}

Here $\alpha$ is determined by the formula

\begin{equation}
\label{3.22} \alpha_{1} =0,\quad \alpha_2 =\frac{d_0 + C_1}{2b_0 -b_3}
\end{equation}

From equation \eqref{3.22} one can find some nonlinear ODE which are useful in physics.

Assuming $a_0=1,\,\,\, a_1=-10, \,\,\, a_2=0, \,\,\, a_3=-10, \,\,\, a_4=0,\,\,\,
a_5=6,\,\,\, b_0=b_1=b_2=b_3=b_4=d_0=d_1=g_0=g_1=d_2=0$ in equation \eqref{3.21} we have
nonlinear ODE in the form

\begin{equation}
\label{3.23v}y_{zzzz} - 10 y^2 y_{zz} -10 yy^2 + 6 y^5 - \alpha C_0 y + \alpha C_1=0
\end{equation}

This is integrable equation and is found from the modified
Korteveg -- de Vries equation of the fifth order \cite{41}

\begin{equation}
\label{3.24v}u_t + \frac\partial{\partial x}\left(u_{xxxx} -10 u^2 u_{xx} -10uu^2 + 6
u^5\right)=0
\end{equation}

Using the travelling wave \eqref{1.2} from \eqref{3.24v} we have equation \eqref{3.23v}.

We can see that equation \eqref{3.23v} is relative to the class of nonlinear ODEs
\eqref{3.21}. Special solution of equation is determined by the formula \eqref{3.2} at
$A_0=0, \,\,\, A_1=1,\,\,\, C_0=6\alpha$ and $C_1=0$.

Assuming $a_0=1,\,\,\, a_1=-5,\,\,\, a_2=5,\,\,\, a_3=-5,\,\,\,
a_4=0,\,\,\, a_5=1,\,\,\,
b_0=b_1=b_2=b_3=b_4=d_0=d_1=d_2=g_0=g_1=0$ in \eqref{3.21} we have
equation in the form \cite{42,43}

\begin{equation}
\label{3.25v}y_{zzzz} - 5 y^2 y_z + 5 y_z y_{zz} - 5yy^2_z + y^5 - \alpha C_0 y + \alpha
C_1 =0
\end{equation}

This is integrable equation too. This one can be found from the Fordy -- Gibbons equation

\begin{equation}
\label{3.26v}u_t + \frac\partial{\partial x}\left(u_{xxxx} -5 u^2 u_{x} + 5u_x u_{xx}
-5uu^2_x + u^5\right)=0
\end{equation}
if we look for solution in the travelling wave \eqref{1.2}.

We get that equation \eqref{3.25v} has special solution \eqref{3.2} at $A_0=0,\,\,\,
A_1=1,\,\,\, C_0=\alpha$ and $C_1=0$.

We can see that a number of integrable equations are abundant in the class of nonlinear
ordinary differential equations \eqref{3.21}.

Assuming $a_1 = 0,\,\, a_2=0,\,\,  a_3 =0,\,\,a_4 =0,\,\,a_5 =0,\,\,b_3=0,\,\, b_2=0,\,\,
b_1 =0,\,\,h_0 = -\frac{C_0}2 + 8 \alpha a_0$ we have

\begin{equation}
\begin{gathered}
\label{3.23}a_{{0}}y_{{{\it zzzz}}}+b_{{0}}y_{{{\it zzz}}}+ \left( -\frac12C_{{0}}+8
\,a_{{0}}\alpha \right) y_{{{\it zz}}}+d_{{0}}y_{{z}}+\\
\\
+ \left( -a_{{4}} +2\,a_{{3}}+6\,a_{{1}}-24\,a_{{0}}-2\,a_{{2}}+a_{{5}} \right)
{y}^{5}+\\
\\
+
 \left( -b_{{3}}-2\,b_{{1}}+b_{{2}}+6\,b_{{0}} \right) {y}^{4}+\\
 \\
 +
 \left( -a_{{5}}\alpha+a_{{4}}\alpha+2\,a_{{2}}\alpha-6\,a_{{1}}\alpha
-2\,a_{{3}}\alpha+24\,a_{{0}}\alpha+C_{{0}} \right) {y}^{3}+\\
\\
+ \left( 2 \,b_{{3}}\alpha-b_{{2}}\alpha-8\,b_{{0}}\alpha+d_{{0}}+2\,b_{{1}} \alpha
\right) {y}^{2}-C_{{0}}\alpha\,y+C_{{1}}\alpha=0
\end{gathered}
\end{equation}

This equation can be found from the nonlinear evolution equation of fifth order that
takes the form

\begin{equation}\begin{gathered}
\label{3.24} u_t + \left(\alpha u + \beta u^2 +\gamma u^3 +\delta u^4\right)u_x+
d_0 u_{xx} + \left(8\alpha a_0 -\frac12 C_0\right) u_{xxx}+\\
\\
 + b_0 u_{xxxx} + a_0 u_{xxxxx}=0
 \end{gathered}
\end{equation}

Some variants of this equation are used at description of
nonlinear waves \cite{44,45,46,47}.

\section{Nonlinear ODEs with exact solutions of the second degree singularity}

Let us find nonlinear ordinary  differential equations with exact
solutions

\begin{equation}
\label{4.1}y(z) =A_0 + A_1 Y + A_2 Y^2
\end{equation}
where $Y(z)$ is a solution of the Riccati equation as well.

Second order class with solutions \eqref{4.1} we found before in
Introduction. Now let us find the third order class of nonlinear
ODEs with solutions \eqref{4.1}.

\textit{\textbf{4.1. Third order ODEs.}} Nonlinear third order
ODEs with solution \eqref{4.1} can have the form

\begin{equation}
\label{4.2}a_0 y_{zzz} + a_1 yy_{z} + b_0 y_{zz} + b_1 y^2 + d_1 y - C_0 y + C_1=0
\end{equation}

Substitution \eqref{4.1} into equation \eqref{4.2} lead to the algebraic equations for
the coefficients of equation \eqref{4.2}. As a result of calculations at $A_2=-12,\,\,
A_1 =12\beta,\,\, A_0=0$ we have

\begin{equation}
\label{4.3}a_1 = a_0
\end{equation}

\begin{equation}
\label{4.4}b_0 = 2 b_1 + 5\beta a_0
\end{equation}

\begin{equation}
\label{4.5}d_0 = (\beta^2 + 8\alpha) a_0 + 10 \beta b_1
\end{equation}

\begin{equation}
\label{4.6}b_1= - \frac{C_0 - \beta^3 a_0 + 4 \alpha\beta a_0}{2(\beta^2 + 8\alpha)}
\end{equation}

\begin{equation}
\label{4.7}\alpha_1 =\frac14 \beta^2,\qquad \alpha_2 =-\frac{C_0}{12\beta a_0}
\end{equation}

Using values for $\alpha$ from \eqref{4.7} we get

\begin{equation}
\label{4.8}b_1^{(1)} =-\frac{C_0}{6\beta^2},\quad b_2 ^{(2)} =\frac12 \beta a_0
\end{equation}

\begin{equation}
\label{4.9}d_0^{(1)} =3\beta^2 a_0 -\frac{5C_0}{3\beta}, \quad d_0^{(2)}=6\beta^2 a_0
-\frac{2C_0}{3\beta}
\end{equation}

\begin{equation}
\label{4.10}b_0^{(1)} =5\beta a_0 -\frac{C_0}{3\beta^2},\quad b_0^{(2)} =6\beta a_0
\end{equation}

\begin{equation}
\label{4.11}C_1^{(1)} =\frac92 \beta^2 C_0, \quad C_1^{(2)} =6\beta^2 C_0 + \frac{C_0^2}
{2\beta a_0}
\end{equation}

We have two nonlinear ODEs with special solutions

\begin{equation}
\begin{gathered}
\label{4.12}a_0 y_{zzz} + a_0 yy_{z} + \left(5\beta a_0 -
\frac{C_0}{3\beta^2}\right)y_{zz} -\\
\\-\frac{C_0 y^2}{6\beta^2} + \left(3\beta^2 a_0 - \frac{5C_0}{3\beta}\right)y_z - C_0 y
+ \frac{9\beta^2}2 C_0=0
\end{gathered}
\end{equation}
and
\begin{equation}
\begin{gathered}
\label{4.13}a_0 y_{zzz} + a_0 yy_{z} + 6 \beta a_0 y_{zz} +\frac12 \beta a_0 y^2+\\
\\
+ \left(6\beta^2 a_0 -\frac{2C_0}{3\beta}\right)y_z  - C_0 y + \frac{C_0^2} {2\beta a_0}
+ 6\beta^2 C_0=0
\end{gathered}
\end{equation}

Solutions of equations \eqref{4.12} and \eqref{4.13} are found by the formula

\begin{equation}
\label{4.14}y(z) =12 \beta Y -12Y^2
\end{equation}
where $Y(z)$ is a solution of the Riccati equations

\begin{equation}
\label{4.15}Y_z = -Y^2 + \frac14 \beta^2
\end{equation}
and
\begin{equation}
\label{4.16}Y_z =-Y^2 - \frac{C_0}{12 \beta a_0}
\end{equation}

Equations \eqref{4.12} and \eqref{4.13} are obtained from nonlinear evolution equation

\begin{equation}
\label{4.17}u_t + \lambda_1 uu_x  + \lambda_2 u_{xxx} +\lambda_3 u_{xxxx} + \lambda_4
(uu_x)_x+ \lambda_5 u_{xx} =0
\end{equation}
if we look for solution of \eqref{4.17} in the form of travelling wave \eqref{1.2}.

Equation \eqref{4.17} was used in of works \cite{48,49,50} for description of nonlinear
waves. Some exact solutions were obtained in papers \cite{28,30}.

\textit{\textbf{4.2. Fourth order ODEs.}} Let us write the general form of the nonlinear
four order ODEs with exact solution \eqref{4.1}. It takes the form

\begin{equation}
\begin{gathered}
\label{4.18}a_0 y_{zzzz} + a_1 yy_{zz} + a_2 y^2_z + a_3 y^3 + b_0 y_{zzz} +b_1 yy_z +\\
\\+
d_0 y_{zz} + d_1 y^2 + h_0 y_z -  \alpha C_0 + C_1 =0
\end{gathered}
\end{equation}

Substitution \eqref{4.1} into equation \eqref{4.12} at $A_2 =1,
\,\, A_1=0,\,\, A_0=0$ gives relations for parameters

\begin{equation}
\label{4.19}a_3 = - 120 a_0 - 4 a_2 - 6 a_1
\end{equation}

\begin{equation}
\label{4.20}b_1 = -12 b_0
\end{equation}

\begin{equation}
\label{4.21}d_1 = 240 \alpha a_0 + 8 \alpha a_2 -6 d_0 + 8\alpha a_1
\end{equation}

\begin{equation}
\label{4.22}h_0= 8\alpha b_0
\end{equation}

\begin{equation}
\label{4.23}d_0=17\alpha a_0 + \frac12 \alpha a_2 -\frac18 C_0 +\frac14 \alpha a_1
\end{equation}

\begin{equation}
\label{4.24}C_1=-18 \alpha^3 a_0 - \alpha^3 a_2 + \frac14 \alpha^2 C_0 - \frac12 \alpha^3
a_1
\end{equation}

We have nonlinear fourth order ODE with exact solution \eqref{4.1} at $A_0=0,\,\, A_1=0$
and $A_2=1$ in the form

\begin{equation}\begin{gathered}
\label{4.25}a_{{0}}y_{{{\it zzzz}}}+b_{{0}}y_{{{\it zzz}}}+ \left( a_{{1}}y+17\,a_
{{0}}\alpha+\frac12a_{{2}}\alpha-\frac18C_{{0}}+1/4\,a_{{1}}\alpha
 \right) y_{{{\it zz}}}+\\
 \\
 +a_{{2}}{y_{{z}}}^{2}+ \left( b_{{1}}y+8\,b_{{0
}}\alpha \right) y_{{z}}+ \left( -120\,a_{{0}}-4\,a_{{2}}-6\,a_{{1}}
 \right)  y^3+\\
 \\
 + \left( \frac34C_{{0}}
+\frac{13}2a_{{1}}\alpha+5\,a_{{2}}\alpha+138\,a_{{0}}\alpha \right) {y}^{
2}-C_{{0}}\alpha\,y+C_{{1}}=0
\end{gathered}\end{equation}

Values $\alpha$ in equation \eqref{2.8} is found from equation \eqref{4.24} at given
parameters $C_1,\,\, a_0,\,\, a_2$ and $C_0$.

Assuming $a_0=1,\,\,\, a_1=-20,\,\,\, a_2=-10,\,\,\, a_3=40,\,\,\, b_0=b_1=d_0=d_1=h_0=0$
in equation \eqref{4.25} we have

\begin{equation}
\label{4.26v}y_{zzzz} -20 yy_{zz} - 10y^2_z + 40 y^3 - \alpha C_0 y + C_1=0
\end{equation}

This is integrable equation. This one is found from the Korteveg
-- de Vries equation of the fifth order \cite{41}.

\begin{equation}
\label{4.27v}u_t + \frac\partial{\partial x} \left(u_{xxxx} -20 uu_{xx} -10u^2_x + 40
u^3\right)=0
\end{equation}

Equation \eqref{4.26v} has special solution \eqref{4.1} at $ A_0=A_1=0, \,\,A_2=1$ if we
take $C_0=56\alpha$ and $C_1=16\alpha^3$.

Assuming $a_0=1,\,\,\, a_0=-18,\,\,\, a_2=-9,\,\,\, a_3=24, \,\,\, b_0=b_1=d_0=d_1=h_0=0$
in equation \eqref{4.25} we get

\begin{equation}
\label{4.28v}y_{zzzz} -18 yy_{zz} + 24 y^3 -9y^2_z -\alpha C_0 y + C_1=0
\end{equation}

This is integrable equation too. This one is found from the
Schwarz -- Kaup -- Kupershmidt equation \cite{51,52}

\begin{equation}
\label{4.29v}u_t+ \frac\partial{\partial x} \left(u_{xxxx} -18uu_{xx} -9u^2_x + 24
u^3\right)=0
\end{equation}

Equation \eqref{4.28v} has special solution \eqref{4.1} at $A_0 = -2\alpha /3,\,\,\,
A_1=0$ and $ A_2=1$ if we take $C_0=0$ and $C_1=-8\alpha^3/9$.

We can see again that our class of nonlinear ODEs contains a number of integrable
equations.

Assuming $a_1=0,\,\, a_2=0$ and $b_1=0$ we have from equation \eqref{4.25}

\begin{equation}\begin{gathered}
\label{4.26}a_{{0}}y_{{{\it zzzz}}}+b_{{0}}y_{{{\it zzz}}}+ \left( -\frac18 C_{{0}}+
17\,a_{{0}}\alpha \right) y_{{{\it zz}}}+8\,b_{{0}}\alpha\,y_{{z}}-120
\,{y}^{3}a_{{0}}+\\
\\
+ \left( \frac34C_{{0}}+138\,a_{{0}}\alpha \right) {y}^{ 2}-C_{{0}}\alpha\,y+C_{{1}}=0
\end{gathered}\end{equation}
 This equation can be obtained from nonlinear evolution equation in the form

\begin{equation}\begin{gathered}
\label{4.27}u_t - 360 a_0 u^2 u_x + \left(\frac32 C_0 + 276 \alpha a_0\right) uu_x+
\\
\\+8 \alpha a_0 u_{xx} + \left(17\alpha a_0 - \frac18 C_0  \right)u_{xxx}
+b_0 u_{xxxx} + a_0 u_{xxxxx} =0
\end{gathered}\end{equation}
if we look for solution using the travelling wave \eqref{1.2}. Equation \eqref{4.27} were
considered before. Exact solutions of this equation were found in works \cite{11,14,
53,54}.

\section{Nonlinear ODEs with solution of the third degree singularity}

Let us find  nonlinear ODEs which have exact solution in the form

\begin{equation}
\begin{gathered}
\label{5.1}y(z) = A_0 + A_1 Y + A_2 Y^2 + A_3 Y^3
\end{gathered}
\end{equation}
where $Y(z)$ is a solution of the Riccati equation too. One can
see that $y(z)$ has the third degree singularity. We have not got
any nonlinear second order ODE in the polynomial form with exact
solutions \eqref{5.1}. However we can find the nonlinear third
order.

\textit{\textbf{5.1. Third order ODEs.}} General form of the nonlinear differential
equation with exact solution \eqref{5.1} is enough simple. It takes the form

\begin{equation}
\begin{gathered}
\label{5.2}a_0 y_{zzz} + a_1 y^2 + b_0 y_{zz} + d_0 y_z - 2C_0 y +
C_1 =0
\end{gathered}
\end{equation}
Using new variables \cite{20} one can write equation \eqref{5.2}
in the form

\begin{equation}
\begin{gathered}
\label{5.3}y_{zzz} + a_1 y^2 + \sigma y_{zz} + y_z - 2C_0 y + C_1
=0
\end{gathered}
\end{equation}

Substituting solution \eqref{5.1} into equation \eqref{5.3} and
taking into account equations \eqref{2.8}, \eqref{2.5} and
\eqref{2.6} we have

\begin{equation}
\begin{gathered}
\label{5.4}A_3 =60,\quad a_1 =1,
\end{gathered}
\end{equation}

\begin{equation}
\begin{gathered}
\label{5.5}A_2 =-\frac{15}{2} \sigma,
\end{gathered}
\end{equation}

\begin{equation}
\begin{gathered}
\label{5.6}A_1=-60\alpha +\frac{30}{19} -\frac{15 \sigma^2}{152}
\end{gathered}
\end{equation}

\begin{equation}
\begin{gathered}
\label{5.7}A_0 =\frac12 C_0 -5\sigma \alpha + \frac7{152} \sigma -\frac{13}{1216}
\sigma^3
\end{gathered}
\end{equation}

We also have two additional equations

\begin{equation}
\begin{gathered}
\label{5.8}-{\frac {75}{19}}\,\alpha\,{\sigma}^{2}+{\frac {1200}{19}}\,\alpha+{ \frac
{1965}{11552}}\,{\sigma}^{4}+{\frac {330}{361}}-{\frac {1305}{
1444}}\,{\sigma}^{2}-480\,{\alpha}^{2}=0
\end{gathered}
\end{equation}
and

\begin{equation}
\begin{gathered}
\label{5.9}\sigma\, \left( \sigma-4 \right)  \left( \sigma+4 \right)  \left( 13\,
{\sigma}^{2}-56+3040\,\alpha \right) =0
\end{gathered}
\end{equation}

From equation \eqref{5.9} we get

\begin{equation}
\begin{gathered}
\label{5.10}\sigma_1=0, \qquad \sigma_{2,3} =\pm4,
\end{gathered}
\end{equation}

\begin{equation}
\begin{gathered}
\label{5.11}\alpha=\frac7{380} - \frac{13}{3040} \sigma^2
\end{gathered}
\end{equation}

Substituting \eqref{5.10} and \eqref{5.11} into equation \eqref{5.8} we obtain

\begin{equation}
\begin{gathered}
\label{5.12}\alpha_1 = - \frac1{76}, \quad \alpha_2 =\frac{11}{76}, \quad \alpha_{3,4}
=\pm\frac14
\end{gathered}
\end{equation}

\begin{equation}
\begin{gathered}
\label{5.13}\sigma_{4,5} =\pm \frac{16}{\sqrt{73}},\quad \sigma_{6,7}=\pm
\frac{12}{\sqrt{47}}
\end{gathered}
\end{equation}

Using $\sigma_{4,5}$ and $\sigma_{6,7}$  we have from equation
\eqref{5.11}

\begin{equation}
\begin{gathered}
\label{5.14}\alpha_5 =\frac1{292},\quad \alpha_6 =\frac1{188}
\end{gathered}
\end{equation}

As this take place constant $C_1$ takes

\begin{equation}
\begin{gathered}
\label{5.15}C_1^{(1)} = C_0^2 + \frac{225}{6859}, \quad C_1^{(2)} = C_0^2
-\frac{2475}{6859}, \quad C_1^{(3)} = C_0^2 -9, \\
\\
 C_1^{(4)} = C_0^2-4,\quad C_1^{(5)} = C_0^2 -\frac{2025}{389017}, \quad C_1^{(6)} =
 C_0^2 - \frac{900}{103823}
\end{gathered}
\end{equation}

As result of calculations we find six equations of equation
\eqref{5.2} with exact solutions.

Assuming $\sigma=0$ and $\alpha=-1/76$ we have equation

\begin{equation}
\begin{gathered}
\label{5.16}y_{{{\it zzz}}}+{y}^{2}+y_{{z}}-2\,yC_{{0}}+{\frac {225}{6859}}+{C_{{0
}}}^{2}=0
\end{gathered}
\end{equation}
with exact solution in the form

\begin{equation}
\begin{gathered}
\label{5.17}y \left( z \right) =C_{{0}}+{\frac {45}{19}}\,Y+60\,{Y}^{3},\,\,\, Y(z) = \pm
\frac{1}{2\sqrt{19}} \,\,\textmd{tg} \left(\pm \frac {z}{2\sqrt{19}} + \varphi_0\right)
\end{gathered}
\end{equation}

At $\sigma=0$ and $\alpha=\alpha_2=11/76$ we obtain

\begin{equation}
\begin{gathered}
\label{5.18}y_{{{\it zzz}}}+{y}^{2}+y_{{z}}-2\,yC_{{0}}-{\frac {2475}{6859}}+{C_{{0
}}}^{2}=0
\end{gathered}
\end{equation}
with exact solution

\begin{equation}
\begin{gathered}
\label{5.19}y \left( z \right) =C_{{0}}-{\frac {135}{19}}\,Y+60\,{Y}^{3},\,\,\, Y(z) =
\pm \frac{\sqrt{11}}{2\sqrt{19}} \,\,\textmd{tanh} \left(\pm \frac {z
\sqrt{11}}{2\sqrt{19}} +\varphi_0 \right)
\end{gathered}
\end{equation}

Assuming $\sigma=4$ and $\alpha=\alpha_3=1/4$ we get

\begin{equation}
\begin{gathered}
\label{5.20}y_{{{\it zzz}}}+{y}^{2}+4\,y_{{{\it zz}}}+y_{{z}}-2\,yC_{{0}}-9+{C_{{0
}}}^{2}=0
\end{gathered}
\end{equation}
with exact solution

\begin{equation}
\begin{gathered}
\label{5.21}y \left( z \right) =\frac92+C_{{0}}-15\,Y-30\,{Y}^{2}+60\,{Y}^{3},\,\,\, Y(z)
= \pm\frac12 \textmd{tanh} \left(\pm \frac z2 +\varphi_0\right)
\end{gathered}
\end{equation}

At $\sigma=4$ and $ \alpha=\alpha_4=-1/4$ we obtain

\begin{equation}
\begin{gathered}
\label{5.22}y_{zzzz} + 4 y_{zzz} + y_z + y^2 - 2 C_0 y + C_0 ^2 -4 =0
\end{gathered}
\end{equation}
with exact solution

\begin{equation}
\begin{gathered}
\label{5.23}y(z) = C_0 -\frac{11}2 + 15 Y - 30 Y^2 + 60 Y^3,\,\,\, Y(z)= \pm\frac12
\,\,\textmd{tg} \left(\pm\frac z2 +\varphi_0\right)
\end{gathered}
\end{equation}

Assuming $\sigma=16/\sqrt{73}$ and $\alpha=\alpha_5=16/\sqrt{73}$ we have

\begin{equation}
\begin{gathered}
\label{5.24}y_{zzz} + \frac{16}{\sqrt{73}} y_{zz} + y_z + y^2 - 2C_0 y + C_0^2=0
-\frac{2025}{389017}
\end{gathered}
\end{equation}
with exact solution

\begin{equation}
\begin{gathered}
\label{5.25}y(z) = C_0 + \frac{485\sqrt{73}}{63948} +\frac{1075}{1168} Y- \frac{120}
{\sqrt{73}} Y^2 + 60 Y^3\\
\\
Y(z)= \pm \frac1{2\sqrt{73}} \,\, \textmd{tanh} \left(\pm \frac{z}{2\sqrt{73}}
+\varphi_0\right)
\end{gathered}
\end{equation}

At $\sigma=12/\sqrt{47}$ and $\alpha=\alpha_6=12/\sqrt{47}$ we get

\begin{equation}
\begin{gathered}
\label{5.26}y_{zzz} + \frac{12}{\sqrt{47}} y_{zz} + y_z + y^2 - 2C_0 y + C_0^2
-\frac{900}{103823}=0
\end{gathered}
\end{equation}
with exact solution

\begin{equation}
\begin{gathered}
\label{5.27}y(z) = C_0 + \frac{45\sqrt{47}}{4418} + \frac{45}{47}Y -
\frac{90\sqrt{47}}{47} Y^2 + 60 Y^3,\\
\\
Y(z) = \pm \frac{1}{2\sqrt{47}} \,\, \textmd{tanh} \left(\pm \frac z{2\sqrt{47}}
+\varphi_0\right)
\end{gathered}
\end{equation}

All these equations can be found from the Kuramoto -- Sivashinsky equation using the
travelling wave \eqref{1.2}. This equation takes the form \cite{55,56,57,58}

\begin{equation}
\begin{gathered}
\label{5.28}u_t + 2 a_1 uu_x + b_0 u_{xxx} + d_0 u_{xx} + a_0 u_{xxxx}=0
\end{gathered}
\end{equation}

 Solutions of this equation were first found in work \cite{11} but they  were
 rediscovered a few times later. At $\sigma=4$ there is periodic solution of equation
 \eqref{5.3} \cite{13}.

 We found that there is only one form of the nonlinear third order differential equation
 with exact solution of the third degree singularity.

\textit{\textbf{5.2. Fourth order ODEs.}} General form of nonlinear fourth order ODEs
with solutions \eqref{5.1} can be presented as the following

\begin{equation}
\begin{gathered}
\label{5.29}a_{{0}}y_{{{\it zzzz}}}+a_{{1}}yy_{{z}}+b_{{0}}y_{{{\it zzz}}}+b_{{1}}
{y}^{2}+d_{{0}}y_{{{\it zz}}}+h_{{0}}y_{{z}}-C_{{0}}\alpha\,y+C_{{1}}=0
\end{gathered}
\end{equation}

For more simple calculations let us first take $A_2 =0$ and $A_3
=120$. We have

\begin{equation}
\begin{gathered}
\label{5.30}a_1 =a_0
\end{gathered}
\end{equation}

\begin{equation}
\begin{gathered}
\label{5.31}b_1 = \frac12 b_0
\end{gathered}
\end{equation}

\begin{equation}
\begin{gathered}
\label{5.32}d_0 = \frac{19}{60} a_0 A_1 + 38 \alpha a_0
\end{gathered}
\end{equation}

\begin{equation}
\begin{gathered}
\label{5.33}h_0 = \frac{19}{60} b_0 A_1 + 38 \alpha b_0 - a_0 A_0
\end{gathered}
\end{equation}

\begin{equation}
\begin{gathered}
\label{5.34}A_{{0}}=\frac75{\frac {a_{{0}}A_{{1}}\alpha}{b_{{0}}}}+ 108{\frac {a_{{0
}}{\alpha}^{2}}{b_{{0}}}}+{\frac {C_{{0}}\alpha}{b_{{0}}}}+\frac{11a_0 A_1^2}{3600\,
b_0}, \,\,b_0\neq0
\end{gathered}
\end{equation}

We also have two values for $\alpha$

\begin{equation}
\begin{gathered}
\label{5.35}\alpha_1^{(1)} = -\frac{A_1} {360},\qquad \alpha_2^{(1)} =-
\frac{11A_1}{1080}
\end{gathered}
\end{equation}

As this takes place we obtain two values of the constant $C_1$

\begin{equation}
\begin{gathered}
\label{5.36}C_{{1}}^{(1)}={\frac {121}{2332800}}\,{\frac {{A_{{1}}}^{2}{C_{{0}}}^{2}}{b_
{{0}}}}+{\frac {11}{43740}}\,{A_{{1}}}^{3}b_{{0}}
\end{gathered}
\end{equation}

\begin{equation}
\begin{gathered}
\label{5.37}C_1^{(2)}={\frac {1}{259200}}\,{\frac
{{A_{{1}}}^{2}{C_{{0}}}^{2}}{b_{{0}}}}+{ \frac {1}{1620}}\,{A_{{1}}}^{3}b_{{0}}
\end{gathered}
\end{equation}

Equation with exact solution
\begin{equation}
\begin{gathered}
\label{5.38}y \left( z \right) =-{\frac {1}{360}}\,{\frac {A_{{1}}C_{{0}}}{b_{{0
}}}}+A_{{1}}Y  +120\, Y ^3
\end{gathered}
\end{equation}
takes the form

\begin{equation}
\begin{gathered}
\label{5.39}a_{{0}}y_{{{\it zzzz}}}+b_{{0}}y_{{{\it zzz}}}+d_{{0}}y_{{{\it zz}}}-{\frac {
\left(360\,a_{{0}}yb_{{0}}+76\,A_{{1}}{b
_{{0}}}^{2}+\,A_{{1}}a_{{0}}C_{{0}} \right) y_{{z}}}{360\,b_{{0}}}}+\\
\\
+\frac12 b_{{0}}{y}^{2}+{\frac {1}{360}}\,A_{{1}}C_{{0}}y+C_{{1}}=0
\end{gathered}
\end{equation}

Where $C_1$ is determined by formulas \eqref{5.36} and \eqref{5.37}.

Assuming $b_0=0$ we have $A_0 \neq 0$ but we find $C_0$ in the form

\begin{equation}
\begin{gathered}
\label{5.40}C_{{0}}=-{\frac {a_{{0}} \left( 5040\,A_{{1}}\alpha
+388800\,{\alpha}^{2}+11\,{A_{{1}}}^{2} \right) }{3600\,\alpha}}
\end{gathered}
\end{equation}

In this case we have

\begin{equation}
\begin{gathered}
\label{5.41}\alpha_1^{(2)} = - \frac{11A}{1080}, \quad \alpha_2^{(2)} = -
\frac{A_1}{360}, \quad \alpha_3^{(2)}= - \frac{A_1}{120}
\end{gathered}
\end{equation}
and three values of constant $C_1$ in the form

\begin{equation}
\begin{gathered}
\label{5.42}C_1^{(1)}=0,\quad C_1^{(2)}=0,\quad C_1^{(3)} = \frac{a_0 A _0 A_1^2}{900}
\end{gathered}
\end{equation}

Equations with exact solution

\begin{equation}
\begin{gathered}
\label{5.43}y \left( z \right) =A_{{0}}+A_{{1}}Y  +120\, Y^{3}
\end{gathered}
\end{equation}

can be presented in the form

\begin{equation}
\begin{gathered}
\label{5.44}a_0 y_{zzzz} + a_0 yy_z - \frac{19a_0 A_1}{270} y_{zz} - a_0 A_0 y_z =0
\end{gathered}
\end{equation}

\begin{equation}
\begin{gathered}
\label{5.45}a_0 y_{zzzz} + a_0 yy_z + \frac{19a_0 A_1}{90} y_{zz} - a_0 A_0 y_z =0
\end{gathered}
\end{equation}

\begin{equation}
\begin{gathered}
\label{5.46}a_0 y_{zzzz} + a_0 yy_z -a_0 A_0 y_z  -\frac{a_0 A_1^2}{900} y + C_1 =0
\end{gathered}
\end{equation}

We have not got any information about physical application of equations \eqref{5.44},
\eqref{5.45} and \eqref{5.46}.

\section{Nonlinear ODEs with exact solutions of the fourth order singularity}

Let us find nonlinear ordinary differential equations which have
exact solutions of the fourth order singularity. These solutions
can be presented by the formula

\begin{equation}
\begin{gathered}
\label{6.1}y \left( z \right) =A_{{0}}+A_{{1}}Y+A_{{2}}{Y}^{2}+A_{{3}}{Y}^{3}+A_{
{4}}{Y}^{4}
\end{gathered}
\end{equation}
where $Y(z)$ satisfies the Riccati equation again.

We can not suggest nonlinear ODEs of the second and third order of the polynomial form
with solution \eqref{6.1}. In this case we can take the fourth order ODE in the form

\begin{equation}
\begin{gathered}
\label{6.2}y_{{{\it zzzz}}}+a_{{1}}{y}^{2}+b_{{0}}y_{{{\it zzz}}}+d_{{0}}y_{{{ \it
zz}}}+e_{{0}}y_{{z}}-C_{{0}}y+C_{{1}}=0
\end{gathered}
\end{equation}

For calculations it is convenient to use first $A_2 =0,\,\,\, A_3
=840\beta /11$ ($\beta$ is new parameter), $A_4= -840n$.
Substituting \eqref{6.1} into equation \eqref{6.2} we have

\begin{equation}
\begin{gathered}
\label{6.3} a_1 =1\quad b_0=\beta,\quad d_{{0}}={\frac {9}{121}}\,{\beta}^{2}+104\,\alpha
\end{gathered}
\end{equation}

\begin{equation}
\begin{gathered}
\label{6.4}e_{{0}}=-{\frac {27}{1331}}\,{\beta}^{3}+{\frac {574}{11}}\,\beta\,
\alpha+{\frac {69}{140}}\,A_{{1}}
\end{gathered}
\end{equation}

\begin{equation}
\begin{gathered}
\label{6.5}A_{{0}}=\frac{C_{{0}}}2-{\frac {90}{121}}\,{\beta}^{2}\alpha+{\frac {81}{29282
}}\,{\beta}^{4}+{\frac {31}{1540}}\,A_{{1}}\beta+816\,{\alpha}^{2}
\end{gathered}
\end{equation}

\begin{equation}
\begin{gathered}
\label{6.6}A_{{1}}=443520\,{\frac {\beta\,{\alpha}^{2}}{-30008\,\alpha+65\,{\beta
}^{2}}}+{\frac {32760}{11}}\,{\frac {{\beta}^{3}\alpha}{-30008\,\alpha
+65\,{\beta}^{2}}}-\\ \\
-{\frac {11340}{1331}}\,{\frac {{\beta}^{5}}{-30008
\,\alpha+65\,{\beta}^{2}}}
\end{gathered}
\end{equation}

We also have equation for $\alpha$ in the form

\begin{equation}
\begin{gathered}
\label{6.7}\left( -12100\,\alpha+27\,{\beta}^{2} \right)  \left( -484\,
\alpha+{\beta}^{2} \right)\times\\
\\
 \times \left( 81\,{\beta}^{6}-696960\,{\beta}^{4}
\alpha+289774672\,{\beta}^{2}{\alpha}^{2}+7029554048\,{\alpha}^{3}
 \right) =0
\end{gathered}
\end{equation}

Assuming

\begin{equation}
\begin{gathered}
\label{6.8}\alpha={\frac {1}{484}}\,{\beta}^{2}
\end{gathered}
\end{equation}

we also have the constant $C_1$ in the form

\begin{equation}
\begin{gathered}
\label{6.9}C_{{1}}=-{\frac {900}{214358881}}\,{\beta}^{8}+\frac{{C_{{0}}}^{2}}{4}
\end{gathered}
\end{equation}

In this case we get equation

\begin{equation}
\begin{gathered}
\label{6.10}y_{{{\it zzzz}}}+\beta\,y_{{{\it zzz}}}+{\frac {35}{121}}\,{\beta}^{2}
y_{{{\it zz}}}+{\frac {13}{1331}}\,{\beta}^{3}y_{{z}}+{y}^{2}-\\
\\
-C_{{0 }}y-{\frac {900}{214358881}}\,{\beta}^{8}+\frac{{C_{{0}}}^{2}}4=0
\end{gathered}
\end{equation}
with exact solution

\begin{equation}
\begin{gathered}
\label{6.11}y \left( z \right) =C_{{0}}+{\frac {45}{29282}}\,{\beta}^{4}-{\frac {
210}{1331}}\,{\beta}^{3}Y+{\frac {840}{11}}\,\beta\,{Y}^{3}-840\,{Y}^{ 4}
\end{gathered}
\end{equation}
where

\begin{equation}
\begin{gathered}
\label{6.12}Y(z) = \pm \frac\beta{22} \textmd{tanh} \left(\pm \frac{\beta z}{22}+
\varphi_0\right)
\end{gathered}
\end{equation}

Assuming $A_1=0,\,\,\, A_2 = 140n,\,\,\, A_3 =0$ and $A_4= - 840$ we have after
calculations

\begin{equation}
\begin{gathered}
\label{6.13}a_1=1,\quad b_0=0, \quad d_{{0}}=-39\,n+104\,\alpha, \quad e_0=0,\\
\\
A_{{0}}=\frac{C_{{0}}}2+816\,{\alpha}^{2}+{\frac {93}{2}}\,{n}^{2}-528\,n \alpha
\end{gathered}
\end{equation}

We get three values for $\alpha$

\begin{equation}
\begin{gathered}
\label{6.14}\alpha_{{1}}=\frac n4, \quad \alpha_{2,3} =\frac14\left(\frac{31}{16}
\pm\frac{i \sqrt{31}}{16}\right)
\end{gathered}
\end{equation}
and the constant $C_1$ in the form

\begin{equation}
\begin{gathered}
\label{6.15}C_{{1}}=-324\,{n}^{4}+\frac{{C_{{0}}}^{2}}4
\end{gathered}
\end{equation}

In this case we obtain equation in the form

\begin{equation}
\begin{gathered}
\label{6.16}y_{{{\it zzzz}}}+{y}^{2}-13\,ny_{{{\it zz}}}-C_{{0}}y-324\,{n}^{4}+\frac{
{C_{{0}}}^{2}}4=0
\end{gathered}
\end{equation}
with exact solution

\begin{equation}
\begin{gathered}
\label{6.17}y \left( z \right) =\frac{C_{{0}}}2-{\frac
{69}{2}}\,{n}^{2}+420\,n{Y}^{2}-840 \,{Y}^{4}
\end{gathered}
\end{equation}
where

\begin{equation}
\begin{gathered}
\label{6.18}Y(z) = \pm \frac{\sqrt{n}}2 \textmd{tanh} \left(\pm \frac{\sqrt{n} z}{2} +
\varphi_0\right)
\end{gathered}
\end{equation}

Exact solutions of equation \eqref{6.16} were found in works
\cite{13,14} and rediscovered in a number of papers later.

\section{Nonlinear ODEs with exact solution of the fifth degree singularity}

Assume that nonlinear ODEs have exact solution of the fifth degree
singularity. Simplest case of this solution takes the form

\begin{equation}
\begin{gathered}
\label{7.1}y \left( z \right) =A_{{0}}+A_{{1}}Y+A_{{2}}{Y}^{2}+A_{{3}}{Y}^{3}+A_{
{4}}{Y}^{4}+A_{{5}}{Y}^{5}
\end{gathered}
\end{equation}

One can see that nonlinearity for equation with solution \eqref{7.1} is $y^2$. We have
term of the tenth degree singularity of this equation. If we want to have the polynomial
form of this equation we have to take

\begin{equation}
\begin{gathered}
\label{7.2}y_{zzzzz} + a_1 y^2 =0
\end{gathered}
\end{equation}

However we can add to equation \eqref{7.2} other terms with lesser
singularity. As a result we have nonlinear ODE in the form

\begin{equation}
\begin{gathered}
\label{7.3}y_{{{\it zzzzz}}}+b_{{0}}y_{{{\it zzzz}}}-d_{{0}}y_{{{\it zzz}}}+e_{{0
}}y_{{{\it zz}}}+h_{{0}}y_{{z}}+a_{{1}}{y}^{2}-C_{{0}}y+C_{{1}}=0
\end{gathered}
\end{equation}

This discussion were given by N.A. Kudryashov in the work \cite{26
} to look for the truncated expansions for nonlinear ODE with the
simplest nonlinearity.

Equation \eqref{7.3} is found from nonlinear evolution equation which takes the form

\begin{equation}
\begin{gathered}
\label{7.4}u_t + 2 a_1 uu_x + h_0 u_{xx} + e_0 u_{xxx} - d_0 u_{xxxx} + b_0 u_{xxxxx} +
u_{xxxxxx} =0
\end{gathered}
\end{equation}

Last years equation \eqref{7.4} was used for the description of the chaos model \cite{59,
60, 61,62} and it is important to have exact solutions of this equation. Let us find
exact solutions of equation \eqref{7.3} at $e_0=0$ and $b_0=0$.

Using new variables

\begin{equation}
\begin{gathered}
\label{7.5}z'=zL^{-1},\quad y' = y B^{-1},\quad L = d_0^{-1/2},\quad B=d_0^{5/2}
a_1^{-1},\\
\\ C'_0=C_0 d_0 ^{-5/2}, \quad C'_1=C_1 d_0 ^{-5/2} B^{-1}
\end{gathered}
\end{equation}

We have equation from \eqref{7.3} in the form

\begin{equation}
\begin{gathered}
\label{7.6}y_{zzzzz} - y_{zzz} + \sigma y_z + \frac12 y^2 - C_0 y + C_1=0
\end{gathered}
\end{equation}
(primes are omitted in equation \eqref{7.6}).

Substituting \eqref{7.1} into equation \eqref{7.6} we obtain

\begin{equation}
\begin{gathered}
\label{7.7}A_5 = 30240, \quad A_4 =0, \quad A_3 =-\frac{2520}{11} - 50400\alpha, \quad
A_2=0,\\
\\
A_1 = \frac{1260}{251} \sigma + 20160 \alpha^2  -\frac{12600}{30371}+ \frac{2520}{11}
\alpha,\quad A_0=C_0
\end{gathered}
\end{equation}

We also have

\begin{equation}
\begin{gathered}
\label{7.8}\sigma={\frac {-92400\,\alpha+10204656\,{\alpha}^{2}+
213811840\,{\alpha}^{3}+2045}{121(9240  \alpha - 79)}}
\end{gathered}
\end{equation}

Taking into account \eqref{7.8} we get six values for $\alpha$

\begin{equation}
\begin{gathered}
\label{7.9}\alpha_1=\frac1{440}, \quad \alpha_2 =\frac5{176}, \quad \alpha_3
=\frac1{220},\\
\\
\alpha_4 =\frac{46031}{52800m}-\frac{557}{52800}- \frac{m}{52800},\\
\\
\alpha_{5,6} =\frac m{105600} -\frac{46031}{105600 m} -\frac{557}{52800}\pm \frac
{i\sqrt{3}}{88} \left(-\frac m{1200}-\frac{46031}{1200 m}\right)
\end{gathered}
\end{equation}

Where

\begin{equation}
\begin{gathered}
\label{7.10}m =\left(113816753 + 1260
\sqrt{8221079733}\right)^{\frac13}
\end{gathered}
\end{equation}

Assuming $\alpha=\alpha_1$ we have equation

\begin{equation}
\begin{gathered}
\label{7.11}y_{{{\it zzzzz}}}-y_{{{\it zzz}}}+{\frac {3259}{12100}}\,y_{{z}}+\frac12{
y}^{2}-C_{{0}}y-{\frac {321489}{322102000}}+\frac12{C_{{0}}}^{2}=0
\end{gathered}
\end{equation}
with exact solution

\begin{equation}
\begin{gathered}
\label{7.12}y \left( z \right) =C_{{0}}+{\frac {189}{121}}\,Y-{\frac {3780}{11}}\,
{Y}^{3}+30240\,{Y}^{5}
\end{gathered}
\end{equation}
where

\begin{equation*}
\begin{gathered}
\label{7.13a}Y(z) = \pm \frac1{2\sqrt{110}} \tanh \left(\pm \frac z{2\sqrt{110}}
+\varphi_0\right)
\end{gathered}
\end{equation*}

At $\alpha=\alpha_2$ we obtain

\begin{equation}
\begin{gathered}
\label{7.13}y_{{{\it zzzzz}}}-y_{{{\it zzz}}}-{\frac {1095}{1936}}\,y_{{z}}+\frac12{y
}^{2}-C_{{0}}y-{\frac {12403125}{82458112}}+\frac12{C_{{0}}}^{2}=0
\end{gathered}
\end{equation}
with exact solution

\begin{equation}
\begin{gathered}
\label{7.14}y \left( z \right) =C_{{0}}+{\frac {4725}{242}}\,Y-{\frac {18270}{11}}
\,{Y}^{3}+30240\,{Y}^{5}
\end{gathered}
\end{equation}
where

\begin{equation}
\label{7.15}Y(z)=\pm \frac{\sqrt{15}}{16\sqrt{11}} \tanh \left(\pm \frac{z\sqrt{5}}{16
\sqrt{11}} +\varphi_0\right)
\end{equation}
Assuming $\alpha=\alpha_3$ we have equation

\begin{equation}
\begin{gathered}
\label{7.16}y_{{{\it zzzzz}}}-y_{{{\it zzz}}}+{\frac {114}{275}}\,y_{{z}}+\frac12{y}^
{2}-2\,C_{{0}}y-{\frac {127008}{20131375}}+2\,{C_{{0}}}^{2}=0
\end{gathered}
\end{equation}

with exact solution in the form

\begin{equation}
\begin{gathered}
\label{7.17}y \left( z \right) =C_{{0}}+{\frac {378}{121}}\,Y  -{ \frac {5040}{11}}
Y^{3}+30240\,
 Y  ^{5},
\end{gathered}
\end{equation}

\begin{equation*}
\begin{gathered}
\label{7.18}Y(z) = \pm \frac{1}{2\sqrt{55}}\, \tanh \left(\pm \frac z{2\sqrt{55}}+
\varphi_0\right)
\end{gathered}
\end{equation*}

We hope these exact solutions will be useful at the study of the
turbulence processes where equation \eqref{7.4} is used.

\section{Conclusion}

Let us emphasize in brief the results of this work. We noted that a lot of nonlinear
differential equations have exact solutions expressed via hyperbolic functions. It is
well known that the base of the hyperbolic functions is the general solution of the
Riccati equation. The widely used tanh method applied for finding exact solutions of many
nonlinear differential equations confirms this idea.

This observation suggested that a lot of special solutions of nonlinear differential
equations can be presented in the form of the general solution of the Riccati equation.
Using this observation we formulated new problem that is to find nonlinear ODEs with
exact solutions. We have found a number of nonlinear differential equations of the
second, third and fourth order which have exact solutions. Exact solutions of these
equations have different singularities and are expressed via general solutions of the
Riccati equation. We also list a number of nonlinear ODEs with exact solutions that are
found from the widely used nonlinear evolution equations.

\section{Acknowledgments}

This work was supported by the International Science and Technology Center under Project
No 1379-2. This material is partially based upon work supported by the Russian Foundation
for Basic Research under Grant No 01-01-00693.


\newpage
\begin{thebibliography}{99}

\bibitem{1} Musette M. and  Conte R., 2003, Physica D, \textbf{181}, 70-79

\bibitem{2} Liu S.K., Fu Z.T., Liu S.D. et al., 2003, Phys. Lett. A., \textbf{309}, 234

\bibitem{3} Yan Z.Y., 2003, Chaos Solitons Fractals, \textbf{15}, 575

\bibitem{4} Yan Z.Y., 2003, Chaos Solitons Fractals, \textbf{15}, 891

\bibitem{5} Elwakil S.A., Ellabany S.K., Zahran M.A. et al., 2002, Phys. Lett. A.,
\textbf{299}, 179

\bibitem{6} Fan E.G., 2001, Nuovo Cimento B, \textbf{116}, 1385

\bibitem{7} Fu Z.T., Liu S.K., Liu S.D. et al., 2001 Phys. Let. A., \textbf{290}, 72

\bibitem{8} Fan E.G., Chao L., 2001, Phys. Lett. A., \textbf{285}, 373

\bibitem{9} Kudryashov N.A., Soukharev M.B., 2001, J. Appl. Math. Mech, \textbf{65}, 855

\bibitem{10} Conte R. (ed) 1999, The Painleve approach ti nonlinear ordinary differential
equations \textit{The Painleve Property. One Century Later (CRM Series in Mathematics
Physics)} (Berlin: Springer)

\bibitem{11} Kudryashov N.A., 1988, Journal of Applied Mathematics and Mekhanics,
\textbf{52},
361-365

\bibitem{12} Conte R. and  Musette M., J. Phys. A., 1989, \textbf{22}, 169-177

\bibitem{13} Kudryashov N.A., 1990, Phys Lett. A., \textbf{147}, 287-291

\bibitem{14} Kudryashov N.A., 1991, Phys Lett. A., \textbf{155}, 269-275

\bibitem{15} Berloff N.G., Howard L.N., 1997, Stud. Appl. Math., \textbf{99}, 1-24

\bibitem{16} Berloff N.G., Howard L.N., 1998, Stud. Appl. Math., \textbf{100}, 195-213

\bibitem{17} Kudryashov N.A., Analytical theary of nonlinear diffeential equations,
Moscow-Igevsk, IKI (2003) 360 p

\bibitem{18} Choudhury S.R., 1997, Phys. Lett. A., \textbf{159}, 311-317

\bibitem{19} Conte R., Musete M., 1992, J. Phys. A.: Math. Gen. \textbf{25}, 5609-5623

\bibitem{20} Kudryashov N.A., and  Zargaryan E.D. 1996, J. Phys. A. Math. and Gen
\textbf{29},
 8067-8077

\bibitem{21} Ince F.L., 1956, Ordinary Differential Equations (New York, Dover)

\bibitem{22} Weiss J.,  Tabor M. and  Carnevalle G., 1983, J. Math. Phys. \textbf{24}, 522-526

\bibitem{23} Pickering A., 1993, J. Phys. A.: Math. Gen., \textbf{26}, 288-300

\bibitem{24} Fan E. G., 2000, Phys Lett. A., \textbf{227}, 212-218

\bibitem{25} Elwakil S.A., Ellabany S.K., Zahran M.A. et al., 2003, Z. Naturforsch,
\textbf{58}, 39-44

\bibitem{26} Kudryashov N.A.,1989, Mathenatical simulation, \textbf{1}, No 9, 151-158 (in Russian)

\bibitem{27} Kudryashov N.A., 1989, Mathematical simulation, \textbf{1}, No 6, 55
(in Russian)

\bibitem{28} Lou S.Y., Huang G.X., Ruan. H.Y., 1991, J. phys. A.: Math. Gen., \textbf{24},
L587-L590

\bibitem{29} Parkes E.J., Duffy B.R., 1996, Comput. Phys. Commun., \textbf{3} 288-300

\bibitem{30} Porubov A., 1993, J. Phys. A.: Math. Gen., \textbf{26}, L797-L800

\bibitem{31} Malfliet W., Hereman W., 1996, Phys. scripta, \textbf{54}, 563-568

\bibitem{32} Malfliet W., 1993, J. Phys. A.: Math. Gen., \textbf{26} L723-L728

\bibitem{33} Conzalez A. and Castellanos A., 1994, Phys. Rev. E., \textbf{49},
2935-2940

\bibitem{34} Jeffrey A. and Xu S., 1989, Wave motion, \textbf{11},
559

\bibitem{35} Berkovich L.M. Factorization and transformation of differential equations, 2002, RHD,
Moscow, 464p. (in Russian)

\bibitem{36} Gudkov V.V. 1995, Journal of numerical mathematics and mathematical physics, \textbf{35},
615-623 (in Russian)

\bibitem{37} Fisher R.A., 1937, Annals of Eugenics, \textbf{7},
355-369

\bibitem{38} Kolmogorov A.N., Petrovskii I.G., Piskunov N.S., 1937, Bulleten MSU, \textbf{1},
1-26

\bibitem{39} Ablowitz M.J. and Zeppetella A., 1979, Bull. Math. Biol., \textbf{41},
835-840

\bibitem{40} Kudryashov N.A., 1994, J. Phys. A.: Math. Gen., \textbf{27}, 2457-2470

\bibitem{41} Ablowittz M.J. and Clarkson P.A., 1991, Solitons, Nonlinear Evolution Equations and inverse
Scattering, Cambridge university press, 516 p.

\bibitem{42} Fordy A.P. and Gibbons J., 1980, Phys. Lett. A., \textbf{160}, 347

\bibitem{43} Hone A.W., 1980, Physica D., \textbf{118}, 1-16

\bibitem{44} Bar D.E., Nepomnyaschy A.A., 1995, Physica D., \textbf{86},
586-602

\bibitem{45} Yamamoto Y., Takizawa E.I., 1981, J. Phys. Soc. Jpn., \textbf{50},
1055-602

\bibitem{46} Kano K., Nakayama T., 1981, J. Phys. Soc. Jpn., \textbf{50},
361

\bibitem{47} Olver V.I., 1984, Hamiltonian and non-Hamiltonian models for
water waves, in Lecture notes in Physics, No. (Springer - Verlag, New York), 273-290

\bibitem{48} Aspe H. and Depassier M.C., 1990, Phys. Rev. A., \textbf{41},
3125

\bibitem{49} Kawahara T and Toh S., 1988, Phys. Fluids, \textbf{31},
2103

\bibitem{50} Garazo A. and Velarde M.G., 1991, Phys. Fluids A, \textbf{3},
2295

\bibitem{51} Weiss J.,1984, J. Math. Phys. \textbf{25}, 13-24

\bibitem{52} Kudryashov N.A., 1999, J. Phys. A.: Math. Gen., \textbf{32}, 999-1013

\bibitem{53} Feng B.F., Kawahara T., 2002, J. bifurcat. Chaos, \textbf{12}, 2393-2407

\bibitem{54} Zemlyanukhin A.I., Mogilevich L.I., 2001 Acoust. Phys., \textbf{47}, 303-307

\bibitem{55} Kuramoto Y. and Tsuzuki T., 1076 Prog. Theor. Phys, \textbf{55}, 356

\bibitem{56} Sivashinsky G.I., 1982, \textbf{4}, 227-235

\bibitem{57} Kawahara T., 1983, Phys. Rev. Lett, \textbf{51}, 381

\bibitem{58} Alfaro C.M., Benguria R.D. and Depassier M.C., 1992, Physica D, \textbf{61},
1-5

\bibitem{59} Beresnev L.A. and  Nikolaevskiy V.N., 1993, Physyca D, \textbf{66}, 1-6

\bibitem{60} Tribelsky M.I. and  Tsuboi K., 1996, Phys. Rev. Lett., \textbf{76}, 1631

\bibitem{61} Xi H.-W.,  Toral R.,  Gunton J.D. and Tribelsky M.I., 2000, Phys. Rev E., \textbf{62}, 17-20

\bibitem{62} Toral R., Xiong G.,  Gunton J.D. and  Xi H.-W., 2003, J.Phys.A. Math. Gen., \textbf{36}, 1323-1335













\end{thebibliography}
\end{document}